\documentclass[floats,apj,twocolumn,nofootinbib]{aastex62}

\usepackage{amssymb}
\usepackage{amsmath}

\usepackage{graphicx}
\usepackage{graphics}
\usepackage[caption=false]{subfig}
\usepackage{dcolumn}
\usepackage{color}
\usepackage{fancyhdr} 
\usepackage{hyperref}

\usepackage[utf8]{inputenc}

\DeclareUnicodeCharacter{00A0}{ }

\usepackage{booktabs}
\usepackage{array}

\makeatletter
\renewcommand*{\fnum@figure}{{\normalfont\bfseries \figurename~\thefigure}}
\renewcommand*{\@caption@fignum@sep}{\textbf{ : }}
\makeatother

\makeatletter
\renewcommand*{\fnum@table}{{\normalfont\bfseries \tablename~\thetable}}
\makeatother

    \newcommand{\be}{\begin{equation}}
  \newcommand{\ee}{\end{equation}}
    \newcommand{\ba}{\begin{align}}
  \newcommand{\ea}{\end{align}}

\usepackage{natbib}
\begin{document}

\title{Periodic Fast Radio Bursts from Young Neutron Stars}

\author{Julian B.~Mu\~noz$^*$
} 
\affiliation{Department of Physics, Harvard University, 17 Oxford St., Cambridge, MA 02138}
\author{Vikram Ravi$^\dagger$
} 
\affiliation{Department of Astronomy, Harvard University, 60 Garden St., Cambridge, MA 02138}
\affiliation{Cahill Center for Astronomy and Astrophysics, MC 249-17, California Institute of Technology, Pasadena CA 91125, USA.}
\author{Abraham Loeb$^\ddagger$
} 
\affiliation{Department of Astronomy, Harvard University, 60 Garden St., Cambridge, MA 02138}

\email{$^*$\tt julianmunoz@fas.harvard.edu}
\email{$^\dagger$\tt vikram@astro.caltech.edu}
\email{$^\ddagger$\tt aloeb@cfa.harvard.edu}

\begin{abstract}
	Fast radio bursts (FRBs) are highly energetic radio pulses from  cosmological origins.
	Despite an abundance of detections, their nature remains elusive.
	At least a subset of FRBs is expected to repeat, as the daily FRB rate surpasses that of any known cataclysmic event, 
	which has been confirmed by observations.
	One of the proposed mechanisms to generate repeating FRBs is supergiant pulses from young and highly spinning NSs, in which case FRBs could inherit the periodicity of their parent NS.
	Here we examine the consequences of such a population of periodic fast radio bursts (PFRBs).	
	We calculate the rate and lifetime of PFRB progenitors, and find that each newly born highly spinning NS has to emit a number $N_{\rm PFRB}\sim 10^2$ of bursts during its active lifetime of $\tau\sim 100$ years, after which it becomes too dim and crosses a PFRB ``death line" analogous to the pulsar one.
	We propose several tests of this hypothesis.
	First, the period of PFRBs would increase over time, and their luminosity would decrease, due to the NS spin-down.
	Second, PFRBs may show modest amounts of rotation measure, given the lack of expelled matter from the pulsar, as opposed to the magnetar-sourced FRBs proposed to explain the first repeater FRB 121102.
	As an example, we study whether the second confirmed repeater (FRB 180814) is a PFRB, given the preference for an inter-pulse separation of 13 ms within its sub-bursts.
	We show that, if confirmed, this period would place FRB 180814 in a different category as FRB 121102.
	We develop tests that would identify---and characterize---the prospective population of PFRBs.
\end{abstract}

\section{Introduction}

Fast radio bursts (FRBs) are mysterious  extragalactic radio transients, lasting for about a millisecond (ms).
Dozens of FRBs have been discovered to date by different telescopes, including Parkes~\citep{Lorimer:2007qn,Thornton:2013iua}, UTMOST~\citep{Farah:2019ahy}, Arecibo~\citep{Spitler:2014fla,Scholz:2016rpt}, ASKAP~\citep{Macquart:2018rsa}, and CHIME~\citep{Amiri:2019qbv}, and a current catalogue can be found in ~\citep{Petroff:2016tcr}.
Yet, our understanding of FRBs is severely lacking (see, for example, \citet{Katz:2018xiu,Popov:2018hkz,Petroff:2019tty,Cordes:2019cmq} for recent reviews).
There are a few requirements that any FRB model must satisfy.
First, the burst source has to be compact enough to produce pulses shorter than a ms, which strongly hints at a compact-object origin, such as neutron stars (NSs)~\citep{Cordes:2015fua,Katz:2017tcd,Popov:2013bia,Connor:2015era,Lyutikov:2016ueh,Falcke:2013xpa,Fuller:2014rza,Totani:2013lia,Abramowicz:2017zbp,Ravi:2014gxa,Popov:2018wei,Metzger:2017wdz,Beloborodov:2017juh}, white dwarfs~\citep{Kashiyama:2013gza}, or black holes~\citep{Barrau:2014yka,Romero:2015nec,Mingarelli:2015bpo}. 
Second, FRBs are very bright and non-thermal at a GHz~\citep{Katz:2013ica}.
Third, FRBs are rather common, with a rate of $\sim 300$ per day above 1 Jy ms~\citep{Bhandari:2017qrj,Amiri:2019qbv}.
The FRB volumetric rate is larger than that of any cataclysmic event that we know, including core-collapse supernovae (CCSN) and NS mergers, pointing to a population of repeating sources, as opposed to one-off events~\citep{Ravi:2019iop}.

This last point was further confirmed by the detection of two repeating FRBs, FRB 121102~\citep{Scholz:2016rpt} and FRB 180814~\citep{Amiri:2019bjk} (R1 and R2 hereafter), as well as additional repeaters recently found by CHIME~\citep{Andersen:2019yex} and ASKAP~\citep{Kumar:2019htf}.
R1, in particular, has been extensively followed-up across different frequencies.
In radio, very-long baseline interferometry (VLBI) observations were able to place R1 in a star-forming galaxy at $z=0.19$, confirming for the first time the cosmological origin of FRBs~\citep{Tendulkar:2017vuq,Marcote:2017wan}.
Additionally, a persistent radio source coincident with R1 was found~\citep{Chatterjee:2017dqg}, hinting at a highly relativistic wind nebula~\citep{Metzger:2017wdz,Beloborodov:2017juh}.
Interestingly, R1 seems distinct from other FRBs, as it shows a rotation measure RM $\sim10^5$ rad m$^{-2}$~\citep{Michilli:2018zec,Gajjar:2018bth}, which is at least three orders of magnitude larger than the RM measured for any other FRB~\citep{Ravi:2016kfj,Masui:2015kmb}, including repeaters~\citep{Andersen:2019yex}.

A concordance model has been developed to explain the data on R1.
In this model, flares from a young magnetar power the FRB emission through a maser synchrotron instability~\citep{Popov:2013bia,Waxman:2017zme,Metzger:2019una},
whereas the persistent radio source is sourced by the pulsar wind nebula (PWN) energized during the early life of the NS~\citep{Katz:2015ltv,Murase:2016sqo,Metzger:2017wdz,Beloborodov:2017juh,Metzger:2019una}.
Additionally, the magnetar flares eject ions out of the star, producing an excess of electrons (versus positrons) in the PWN, 
thus explaining the large RM~\citep{Metzger:2017wdz,Margalit:2018dlt}.

While this magnetar model succesfully explains most data regarding R1, it is far from the only mechanism able to generate FRBs (albeit is worth noting that all mechanisms are hypothetical at this point, 
and are yet to be confirmed).
In particular, it has been shown that supergiant pulses (SGPs; analogous to those of the Crab pulsar~\citep{Cordes:2003im,Mickaliger:2012me,CrabHankins}) from young highly spinning NSs are bright enough to produce observable FRBs~\citep{Cordes:2015fua,Pen:2015ema,Lyutikov:2016ueh,Connor:2015era}. 
In this case the energy reservoir is the rotation of the young NS, which is tapped through its spin-down.
We build upon these previous references, and argue that this would give rise to a qualitatively different population of FRBs from the magnetar model, which we label Periodic Fast Radio Bursts (PFRBs).
Given the current FRB rate from~\citep{Ravi:2019iop}, we estimate that, if an $\mathcal O(1)$ fraction of FRBs are periodic, each PFRB would have to repeat $N_{\rm rep}\sim 200$ times during its lifetime of $\tau\sim 100$ years in order to explain their daily rate.
This is in line with the statistics of SGPs in the Crab pulsar, as shown in~\citeauthor{Cordes:2015fua} (2016; CW16 hereafter).
Moreover, we develop several predictions of this model that can be tested with upcoming FRB data, which will separate a prospective PFRB population from magnetar-induced FRBs.

As an example, we entertain the possibility that R2 is a PFRB.
Intriguingly, R2 appears to show a 13 ms period within its sub-pulses. 
We show that, if confirmed, such a period would imply that R2 is a rotationally driven PFRB, dissimilar from R1.
We then predict that, over the next decade, the typical fluence of R2 would drop by a factor of $\sim$2, and its period would increase to $\sim 16$ ms.
Additionally, polarization data of R2 should show a small rotation measure, bounded by $\rm |RM| \lesssim 80$ rad m$^{-2}$.
These three key predictions, qualitatively common to all PFRB candidates, are easily testable with data.

The outline of this paper is as follows.
In Sec.~\ref{sec:Energy} we review the energetics and rates of FRBs, and how to explain them with spinning neutron stars.
We lay our predictions in Sec.~\ref{sec:Predictions}, which we discuss in Sec.~\ref{sec:Conclusions} before concluding.

\section{Energetics and Rate}
\label{sec:Energy}

\subsection{PFRB Energetics}

We begin by reviewing the energetics of how supergiant pulses (SGPs) can boost the radio emission of a young NS, producing FRBs.
This section draws heavily from CW16, as well as~\citep{Pen:2015ema,Lyutikov:2016ueh,Connor:2015era}, so the  reader familiar with those references might want to skip ahead.

The energy requirements of FRBs are fairly strict, especially if they originate at cosmological distances, as proven by the three localized FRBs to date~\citep{Chatterjee:2017dqg,Ravi:2019alc,Bannister:2019iju} (although a subpopulation of PFRBs could lurk closer to us~\citep[CW16,][]{Connor:2015era}.
From the $\sim$ Jy ms fluences observed, at half a Gpc distances, we require radio luminosities $L_r \geq 10^{40}$ erg s$^{-1}$ during the FRB.
While this is far too large for all known Galactic NSs (even for the Crab pulsar during a SGP), a young highly spinning NS can reach such luminosities.
Given a NS with period $P$ and period derivative $\dot P$, the rate at which it loses its rotational energy (its spin-down luminosity) is $L_{\rm sd} \propto P^{-3} \dot P$.
Only a part of this energy is emitted in radio waves, though, which we parametrize through a radio effiency $\epsilon_r<1$. 
Then, the spin-down luminosity required to emit enough radio energy as a typical FRB is
\be
L_{\rm sd} \geq 10^{41}\,{\rm erg\, s^{-1}} \times \left(\dfrac{\epsilon_r}{10^{-2}}\right)^{-1} \left(\dfrac{f_b}{10^{-1}}\right),
\label{eq:LSD}
\ee
where $f_b<1$ is the beaming factor of the NS.

Here, and throughout this work, we set a radio effiency $\epsilon_r = 10^{-2}$, which albeit typical of old pulsars, is anomalously large for a young neutron star~\citep{Szary:2014dia}.
Nonetheless, SGPs as those seen in the Crab pulsar~\citep{Cordes:2003im}, can reach such large instantaneous radio efficiencies~\citep[CW16,][]{Lyutikov:2016ueh}.
Interestingly, this could explain the episodic nature of some FRBs, as only during periods of intense radio activity (large $\epsilon_r$) would they be observable in the radio.

\begin{figure*}[hbtp!]
	\centering
	\includegraphics[width=0.72\textwidth]{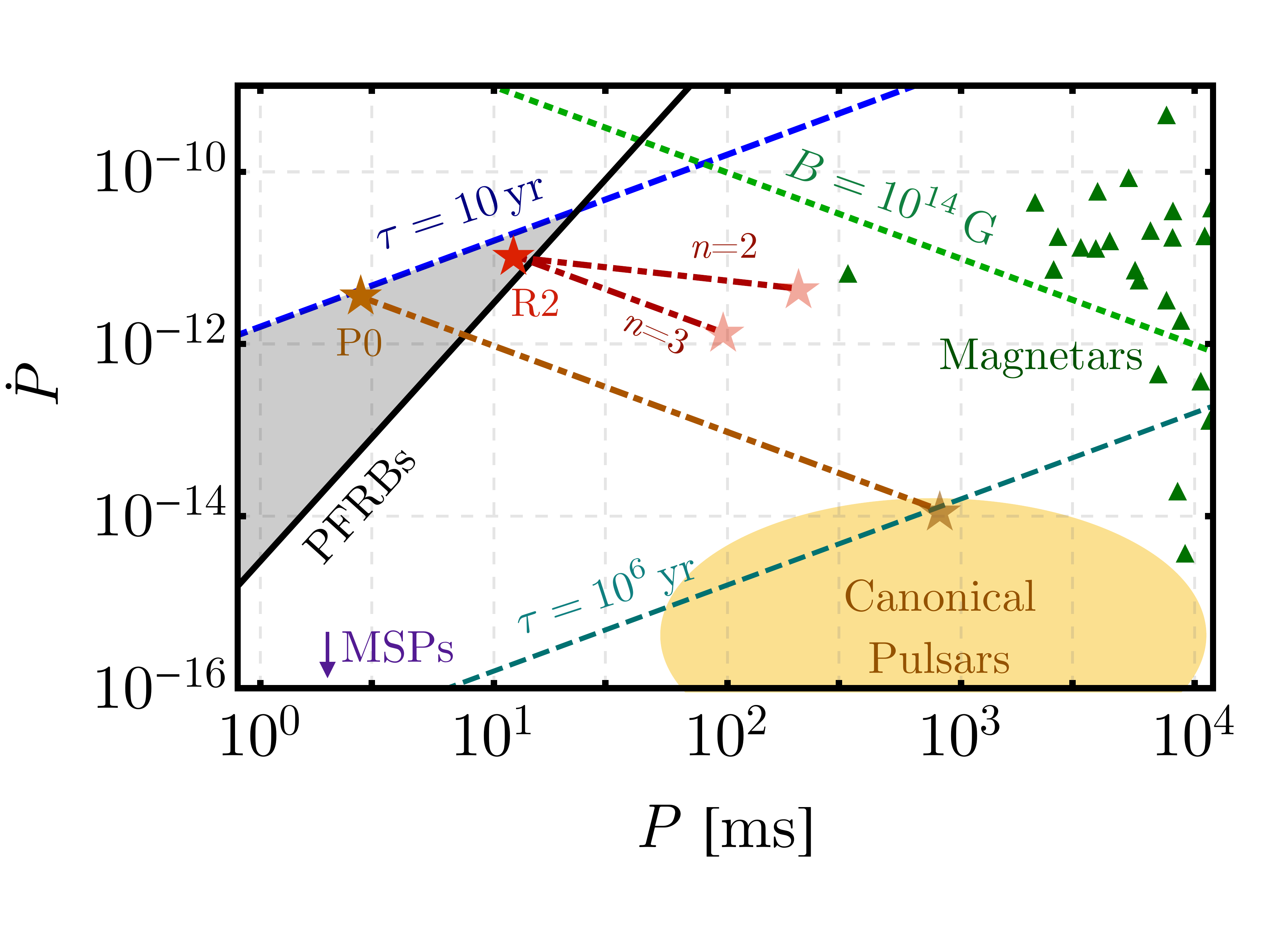}
	\caption{
		Period ($P$, in ms) and period derivative ($\dot P$) diagram for pulsars and PFRBs.
		We show the known magnetars (from the McGill catalogue~\citep{Olausen:2013bpa}) as green triangles, and the canonical pulsars as a yellow blob. 
		The known millisecond pulsars (MSPs) lie below the reach of this plot ($\dot P\approx 10^{-20}$).
		NSs with magnetic fields above $B = 10^{14}$ G (shown as the dotted green line) can produce magnetically driven FRBs through flares.
		Here, instead, we focus on the less-magnetic NSs in the gray shaded area, which can emit periodic fast radio bursts (PFRBs) as supergiant pulses, powered through their spin-down. 
		Below the black line, PFRBs would have a spin-down luminosity lower than $10^{41}$ erg s$^{-1}$, crossing the FRB death line and becoming invisible (at half a Gpc).
		Above the blue-dashed line pulsars would have a characteristic age $\tau<10$ years, so their SNRs would be opaque to FRBs.
		We show, as a red star, the possible location for the second repeater R2 (FRB 180814), if the period of 13 ms is confirmed; and the red dash-dotted lines represent its evolution over $10^3$ years assuming braking indices of $n=2$ and $3$, respectively.
		We also show, as a brown star, a hypothetical young highly spinning NS, which we have dubbed P0, as a possible PFRB source, which would evolve to become a canonical pulsar (at around $\tau=10^6$ years, which is marked by the dashed cyan line).
	}	
	\label{fig:Ppdot}
\end{figure*}

We show in Fig.~\ref{fig:Ppdot} the part of the $P-\dot P$ plane that satisfies the energy requirement of Eq.~\eqref{eq:LSD}, and can thus power PFRBs through their rotational energy alone.
These are highly spinning $(P\lesssim20$ ms) young NSs.
Interestingly, if we require the PFRB progenitors to be older than $10$ years, so that the supernova remnant (SNR) is not opaque to GHz radio emission~\citep{Metzger:2017wdz,Bietenholz:2019urr}, while still being rotationally powered, we find that the allowed part of the $P-\dot P$ plane has small $B$ fields.
Indeed, these would be below the magnetar line, at a surface magnetic field of $B = 10^{14}$ G, where the $B$ field can break the NS crust and carry material out~\citep{Duncan:1992hi}.
Such high fields are necessary to produce the synchrotron maser mechanism from Refs.~\citep{Popov:2013bia,Waxman:2017zme,Metzger:2017wdz,Beloborodov:2017juh,Metzger:2019una,Metzger:2017wdz,Margalit:2018dlt},
as well as the RM observed in R1.
The young NSs that we are considering have, on the other hand,  few-ms periods at birth, but relatively small B fields.
While challenging to confirm~\citep{Kaspi:2002pu,Noutsos:2013ce},  such rapid rotation is physically allowed, as the breakup period of a typical NS with an 11 km radius and 1.5 $M_\odot$ mass is $P \approx 0.5$ ms, and in fact ms periods have been argued to be required to generate the large $B$ fields seen in magnetars~\citep{Duncan:1992hi}.
PFRB sources can be ``frustrated magnetars", where the $B$ field did not grow exponentially.

Thus, the magnetosphere of the NS is the most likely source of coherent radio emission for PFRBs, which will result in dissimilar predictions from the magnetar-powered model of FRBs, as we will explore below.

\subsection{PFRB Rates}

Let us next extrapolate the known statistics of the Crab pulsar to estimate how often PFRBs would happen.
In addition to the giant pulses (GPs) commonly observed in the Crab pulsar, and its twin PSR B0540-69~\citep{Johnston:2003xa,Mickaliger:2012me}, there is a tail of high-luminosity events, typically denoted as supergiant pulses (SGPs)~\citep{Cordes:2003im}.
Following CW16, we join the beaming and efficiency parameters into a single $\zeta_r = \epsilon_r/f_b$, which ought to be $\zeta_r\approx 0.1$ to power PFRBs (although $\zeta_r\sim 10^{-2}$ is enough to produce ``local" FRBs up to $\sim 100$ Mpc, as suggested in CW16 and \citep{Connor:2016imq}.
For the Crab it is estimated that there is a pulse with $\overline{\zeta_r}=0.002$ once per hour~\citep{Crossley:2010ba}, yielding a rate $\overline{\mathcal R} = 10^{-5}$ per cycle, given the Crab period.
While the Crab GPs appear to follow Poissonian statistics, we do not know whether the same is true for SGPs (and thus of PFRBs)~\citep{Connor:2016imq}.
We adopt the approach of CW16 and assume that the rate $\mathcal R$ of events with efficiency $\zeta_r$ is given by a power-law,
\be
\mathcal R(\zeta_r) = \overline{\mathcal R} \times \left(\dfrac{\zeta_r}{\overline{\zeta_r}}\right)^{-\beta},
\ee
where $\beta$ is the power-law index, which is fairly unconstrained for SGPs (although for the more-common GPs this index is on the range of $\beta\sim2-3$).
We set $\beta=2.5$, which reproduces the observation of a 2 MJy SGP from~\citep{CrabHankins} (corresponding to $\zeta_r=0.02$~\citep[CW16,][]{Lyutikov:2016ueh}) roughly once per 20 days (although this particular SGP is at 9 GHz, instead of 430 MHz as the more common SGPs).
We see that by extrapolating this rate to higher $\zeta_r$ (as no cutoff has been found~\citep{Cordes:2003im}), we obtain a rate of one FRB ($\zeta_r=0.1$) per $2\times 10^9$ pulses.
This rate is altered to one FRB per $\{3\times 10^8, 10^{10}\}$ pulses for $\beta=\{2,3\}$, showing a strong dependence on this unknown parameter.
We estimate that a NS born with $P=3$ ms and $\dot P=10^{-11.5}$ will go over $N_{\rm cycle} = \int dt P^{-1}(t) \approx 5 \times 10^{11}$ cycles during its lifetime ($\tau=100$ years), so it will produce $\sim 10^2$ PFRBs (bursts with  $\zeta_r \approx 0.1$).
Here we have conservatively estimated the variability of the young NS using that of the Crab as a reference.
Recent work suggests that Crab GPs are correlated with glitches~\citep{Kazantsev:2019vfq}, so it is likely that younger, less stable pulsars would show even larger variability.

\subsection{Comparison with Observations}

We now compare the SGP rate and energetics outlined above, with FRB observations.
Recent estimates have found a cosmic FRB rate~\citep{Amiri:2019qbv,Ravi:2019iop}
\be
\mathcal R_{\rm FRB} \gtrsim 10^5\,f_b^{-1} \,\rm Gpc^{-3}\, yr^{-1}.
\ee
Thus, FRBs are more abundant than any known cataclysmic event, showing that there has to exist a population of repeating FRBs~\citep{Ravi:2019iop}.
In~\citep{Taylor:2014rlo} it was estimated that the rate of CCSN  (the progenitors of NSs) in the local Universe ($z\lesssim 0.1$) is
\be
\mathcal R_{\rm CCSN} \approx 1.1 \times 10^5 \,\rm Gpc^{-3}\, yr^{-1},
\ee
so if a fraction $f_{\rm CCSN}$ of CCSN become FRB emitters and, as in Eq.~\eqref{eq:LSD}, assume a beaming factor $f_b=0.1$ for those~CW16, we  estimate the number of times that each FRB has to repeat (above threshold) simply as
\be
N_{\rm rep} \gtrsim \dfrac{\mathcal R_{\rm FRB}}{\mathcal R_{\rm CCSN} f_{\rm CCSN} f_b} \gtrsim 10^2 \left(\dfrac{f_b}{0.1}\right)^{-1} \left(\dfrac{f_{\rm CCSN}}{0.1}\right)^{-1},
\label{eq:Nrep}
\ee
assuming that all FRBs are powered by the rotation of young NSs (i.e., PFRBs).
Thus, each PFRB progenitor ought to emit at least a hundred bright SGPs during its lifetime to account for the entire FRB population.
This is in agreement with the statistics of the SGPs estimated above.
Given the steep decline in the spin-down luminosity of young NSs ($L_{\rm sd} \propto t^{-2}$), the typical lifetime that a NS can power an FRBs is a rather short $\sim 30-100$ yrs (depending on the initial $P$ and $\dot P$), an order of magnitude below that of the magnetar model ($300-1000$ yrs)~\citep{Nicholl:2017slv}.
Note that here we have set a fiducial value of the fraction $f_{\rm CCSN}=10\%$ of CCSN that result in the young highly spinning NSs, as it is comparable the birth rate of magnetars, although this fraction is highly uncertain and difficult to reconstruct from data~\citep{FaucherGiguere:2005ny,Johnston:2017wgm}.

Additionally, assuming that all FRBs observed thus far are repeaters, we can estimate their comoving number density to be
\be
n_{\rm PFRB} \equiv \mathcal R_{\rm FRB} \tau_{\mathrm{PFRB}} \approx 3\times 10^7\,\left(\dfrac{f_b}{0.1}\right)^{-1} \,\rm Gpc^{-3}
\ee
where we have chosen a representative lifetime of $\tau_{\mathrm{PFRB}}\approx 30$ yrs.
Thus, the odds of having a PFRB source within the closest Mpc to us---and pointing towards Earth---are $\gtrsim 0.3\%$ (depending on how overdense the local environment is), and any burst emitted by such a source would be detectable with low-cost antennae~\citep{Maoz:2017qpw}.
The PFRB source density estimate above is between that of the magnetar model, which has $n_{\rm FRB} \sim 10^4\,\rm Gpc^{-3}$~\citep{Nicholl:2017slv}, and that of the ``wandering beams" model (where FRBs are emitted from regular pulsars with a beaming factor $f_b\sim 10^{-8}$)~\citep{Katz:2016zxi}, which requires $n_{\rm FRB} \approx 10^{15} \,\rm Gpc^{-3}$.

\section{Predictions}
\label{sec:Predictions}

The PFRB model outlined above has a series of unique predictions, different from those of other mechanisms (including magnetars), which we now outline.
We emphasize that, as is the case for gamma ray bursts, multiple FRB populations can coexist, and might in fact be preferred by current data~\citep{Palaniswamy:2017aze,James:2019tum,Caleb:2019szc}. 
Separating and characterizing each FRB population will be an important step towards better understanding this phenomenon.

For illustration purposes, throughout this section we will refer to a hypothetical young NS with $P=3$ ms and $\dot P=10^{-11.5}$ as P0, shown as a star in Fig.~\ref{fig:Ppdot}, which would be an example of a PFRB emitter.
The inferred surface $B$ field of this source would be $B=3\times 10^{12}$ G, and its characteristic age $\tau=15$ yr.
Unless otherwise stated, we will assume a standard braking index $n=3$, ignoring $B$-field decay during the short period of the NS life that we study~\citep{Johnston:2017wgm}.
Our predictions belong to two categories, those related with the spin-down nature of PFRBs in our model, and those related to their lack of strong $B$ fields.
These are:

$\bullet$ PFRB slow down. 
In the periodic model of FRBs every source is a highly spinning NS with some period $P$ and derivative $\dot P$.
Given the high energetics required to power FRBs, as given by Eq.~\eqref{eq:LSD}, all candidate PFRBs will have short periods ($P\lesssim 20$ ms)
which slow down relatively quickly. 
For our example P0, the period will change by 30\% (from 3 to 4 ms) over a decade.
Such a change is potentially observable via continuous monitoring, once periodicity is established. 

$\bullet$ FRB dimming over time. 
The spin-down timescale for the young NSs that we are considering is typically short ($t_{\rm sd} \lesssim 10\,{\rm yrs}$)~\citep{Kashiyama:2017ehl}, so the spin-down luminosity will decrease as $t^{-2}$ during the entire PFRB life cycle.
This scaling would be imprinted onto the FRB fluences, yielding dimmer FRBs from older sources (which would thus appear more rare for a fixed flux threshold). 
In fact, it is possible that some ``one-off" FRBs are anomalously bright pulses from otherwise dim PFRBs, in anology with the RRAT case for pulsars~\citep{McLaughlin:2005eq}.

$\bullet$  A closer population of FRBs. Farther sources would have to be younger to be bright enough to be detectable. Nonetheless, the SNR around the NSs does not allow FRBs to escape for sources younger than $\tau\approx 10$ yrs. 
Thus, there exists a natural ``horizon" for these FRBs, previously estimated at $\sim$100 Mpc~\citep[CW16,][]{Connor:2015era}. This horizon can be expanded for NSs born with ms periods, albeit not farther than a few Gpc (for our assumed radio efficiencies and beaming factors). For instance, our example P0 with $P=3$ ms and $\dot P=10^{-11.5}$ will be observable with Jy ms fluence up to $z=0.5$ (2 Gpc comoving distance).
An object that far would, however, quickly spin down and become unobservably dim. 
Thus, in this model older FRBs would tend to reside closer to us.

$\bullet$  Given the  typical short lifetimes of PFRBs, we expect the SNR around them to make a non-negligible contribution to their DM~\citep{Piro:2016aac,Yang:2017vtd}. For a typical ejecta mass of $M_{\rm ej}=10\,M_\odot$, an ionization fraction of 10\%~\citep{Metzger:2017wdz}, and a velocity of $v_{\rm SNR}=10^9$ cm s$^{-1}$, we find that the SNR contributes a dispersion measure component of
\be
{\rm DM_{\rm SNR}} \approx  30  \,{\rm pc\,cm^{-3}} \times \left (\dfrac{\tau}{30 \,\rm yrs}\right)^{-2},
\label{eq:SNRDM}
\ee
for an object of age $\tau$, which varies at a rate of
\be
\dfrac{d{\rm DM_{\rm SNR}}}{dt} \approx  -2\, {\rm pc\,cm^{-3}\,yr^{-1}}\times \left (\dfrac{\tau}{30 \,\rm yrs}\right)^{-3},
\label{eq:SNRDMdot}
\ee
which is potentially observable~\citep{Hessels:2018mvq}.
This would make the DM of these sources quickly varying, unlike other FRB models, although it can be negligible if the ejecta was less massive~\citep{Kashiyama:2017ehl}.

$\bullet$ Small RM.  The young NSs that we consider do not have surface magnetic fields large enough to eject ions from their crust, as opposed to magnetars~\citep{Duncan:1992hi}. 
Thus, the PWN around the NS has equal amounts of electrons and positrons (without the presence of ions to compensate), and cannot generate large RMs as observed for R1~\citep{Michilli:2018zec,Gajjar:2018bth}.
The SNR, on the other hand, can potentially produce a non-negligible RM if the $B$ field is aligned with the line of sight, and does not show many reversals.
We follow the estimate in~\citep{Connor:2015era}, where assuming $B_{||} \approx 1\, \mu G$ (and no reversals) we find
\be
|{\rm RM_{\rm SNR}}| \approx 24 {\,\rm rad\,m^{-2}} \times \left(\dfrac{\rm DM_{\rm SNR}} {30  \,{\rm pc\,cm^{-3}} }\right),
\label{eq:SNRRM}
\ee
which will quickly decay as the SNR expands, and can never reach the $\rm |RM|\sim 10^5$ rad m$^{-2}$ values of R1.
It can, however, approach $\rm |RM|\sim 10^2$ rad m$^{-2}$, as recently reported for the repeater FRB 180916.J0158+65~\citep{Andersen:2019yex}.

\subsection*{FRB 180814}

Next we entertain the possibility that the recently detected FRB 180814 (R2;~\citep{Amiri:2019bjk}) is an example of a PFRB.
Our main motivation is that the subpulses detected on September 17$^{\rm th}$ and October 28$^{\rm th}$ appear to be preferentially separated by 13 ms.
If this were indeed the intrinsic period of a NS sourcing R2, this source would be the first periodic FRB detected and, as we will show, would be best explained by the PFRB model (as opposed to a flaring magnetar).
Additionally, we can exemplify the predictions on PFRBs outlined above for the specific case of R2.

First, from an energetics perspective, we can place R2 in the $P-\dot P$ plane by requiring enough spin-down luminosity to power a PFRB out to its maximum redshift $z=0.1$ (400 Mpc), and an age $\tau \geq 10$ years, which we do in Fig.~\ref{fig:Ppdot}.
Note that this maximum distance to R2 is significantly smaller than for R1, as is expected of PFRBs~\citep[CW16,][]{Connor:2015era}.
A NS with this $P$ and $\dot P$ has a relatively low surface magnetic field ($B\approx 10^{13}$ G), making it an ideal PFRB candidate.
Moreover, given the required $\dot P \sim 10^{-11}$, we predict that the period of R2 will increase by $2\%$ per year, reaching a value of 16 ms in a decade, at which point the spin-down power will have decreased by a factor of 2 with respect to its value today.

Second, we can constrain the age of R2 to be above 15 years by demanding that the SNR contribution to its DM, from Eq.~\eqref{eq:SNRDM}, is smaller than its observed extragalactic component ($\rm DM_{\rm EG}\approx100$ pc cm$^{-3}$~\citep{Amiri:2019bjk}).
Additionally, using the full data in~\citep{Amiri:2019bjk} we fit for a time derivative of this quantity, and find $d\rm DM_{\rm EG}/{\it dt} = 15 \pm 20$ pc cm$^{-3}$ yr$^{-1}$, consistent with zero.
By comparing this bound with Eq.~\eqref{eq:SNRDMdot} we also find that $\tau\geq15$ years.
For the predicted values of $P-\dot P$, the characteristic age of the progenitor of R2 is 22 years, in agreement with the constraints above, as well as with with the required $\tau\geq10$ years for the remnant to be transparent to radio~\citep{Metzger:2017wdz}.

Finally, all detections of R2 lack polarization information, so no conclusion about its RM has been reached thus far.
In the PFRB model we expect only a modest RM from the SNR, which is bounded by $|{\rm RM} |\leq 80$ rad m$^{-2}$, found by using $\rm DM_{EG}$ as the SNR contribution in Eq.~\eqref{eq:SNRRM}.

\section{Discussion and Conclusions}
\label{sec:Conclusions}

In this paper we have studied the predictions of a model in which FRBs are powered by rotational energy from young NSs, and are thus periodic (PFRBs).
While we have focused on the radio burst itself, emission is expected in other bands.
For instance, higher-energy photons will energize a PWN around the NS, which can give rise to a persistent radio source, akin to that of R1.
This mechanism has been extensively studied, and we refer the reader to Refs.~\citep{Murase:2016sqo,Kashiyama:2017ehl,Margalit:2018bje,Beloborodov:2017juh,Wang:2019oxv} for more details.
Moreover, the PFRB progenitors can be fairly bright in X-rays~\citep{Perna:2007ww,Popov:2016jcu,Lyutikov:2016qio}, although these do not escape the opaque SNR for a few hundred years, hindering detection~\citep{Metzger:2017wdz,Bhirombhakdi:2018hil}.
A more-direct avenue are gamma-rays, although current limits do not constrain this model~\citep{Cunningham:2019xbu}.

While we have shown that energetically it is possible that NSs emit PFRBs, we have not addressed the actual emission mechanism.
For instance, any successful model for repeating FRBs has to explain the downward drift in frequency commonly observed.
This drift could be intrinsic to the emission process, as for instance both curvature radiation~\citep{Katz:2018afd,Wang:2019bpi,Lu:2017prv,Kumar:2017yiq} and a mechanism analogous to solar flares~\citep{Fletcher:2011ui} are expected to produce this drift, whereas propagation effects (such as plasma lensing) typically produce drifts in both directions~\citep{Cordes:2017eug}.
Given the uncertainties surrounding the radio emission mechanism of pulsars in general, and of the Crab SGPs in particular, we leave the task of modeling the PFRB emission for future work.

In the PFRB scenario the episodic nature of FRBs can be explained by the stochasticity of SGP emission (as only a few times per year there are SGPs above threshold).
In addition, starquakes or ``storms" on NS magnetospheres~\citep{Katz:2017tcd,Wang:2017agh} might produce sporadic releases of energy.
On the other hand, a continually emitting source might appear episodic if observed through a dense---but porous---SNR, as the observer and source move. For instance, over/underdensities in the SNR of size $\sim 100$ km can explain the $\sim 0.1$ s duration of some of the bursts in R2.
We note, in addition, that the magnetar model can potentially generate periodic FRBs under specific conditions~\citep{Beloborodov:2017juh}.

As an application, we note that any cosmological source of periodic bursts can act as a ``standard clock", allowing us to measure the cosmic expansion rate through the time dependence of redshifts~\citep{Loeb:1998bu}.
Unfortunately, this effect induces a diminute $\dot P \sim H_0 P \sim 10^{-20}$ on PFRBs, orders of magnitude below the value necessary to power them.
While challenging to observe, this effect would add to the cosmological power of FRBs~\citep{McQuinn:2013tmc,Munoz:2016tmg,Walters:2017afr,Munoz:2018mll,Madhavacheril:2019buy,Ravi:2019acz}, 
especially if some are strongly lensed~\citep{Dai:2017twh,Li:2017mek,Wagner:2018fvv}.

In conclusion, in this work we have studied the possibility that a population of FRBs are periodic repeaters, which we have dubbed PFRBs.
An example might be the recently discovered FRB 180814, which appears to show a 13-ms period.
We have put forth a series of predictions, which can test this hypothesis within the next decade, shedding light onto the origin of the mysterious fast radio bursts.

We are thankful to Jim Cordes and Shami Chatterjee for discussions,
and the owners of the FRB~\footnote{\url{http://frbcat.org/}} and magnetar~\footnote{http://www.physics.mcgill.ca/~pulsar/magnetar/main.html} catalogues.
JBM is funded by a Department of Energy (DOE) grant de-sc0019018.
AL is supported in part by grants from the JTF to the BHI, and by the Breakthrough Prize Foundation.

\bibliography{Rotational_FRBs}

\end{document}